\documentclass[floatfix,
 reprint,
 amsmath,amssymb,
 aps,
prl,
]{revtex4-2}
\usepackage[utf8]{inputenc}
\usepackage{dcolumn}
\usepackage{wasysym}
\usepackage[caption=false]{subfig}
\usepackage{graphicx}
\usepackage{here}
\usepackage{epstopdf}
\usepackage{array}
\usepackage{amsmath}
\usepackage{physics}
\usepackage{verbatim}
\usepackage{amsfonts,amssymb,amscd,mathtools}
\usepackage{mathalfa}
\usepackage{textcomp}
\usepackage{quantikz}
\usepackage{amsthm}
\usepackage{tabularx}
\usepackage{stmaryrd}
\usepackage{enumerate}
\usepackage{wasysym}
\usepackage{braket}
\usepackage{mathrsfs}
\usepackage{microtype}

\usepackage[dvipsnames]{xcolor}

\usepackage{tikz}
\usepackage[lmargin=.7in,rmargin=.7in,tmargin=.7in,bmargin=1in]{geometry}
\usepackage{newtxtext,newtxmath}
\usepackage{algorithm}
\usepackage{algorithmicx}
\usepackage{algpseudocode}


\usepackage{booktabs}
\theoremstyle{plain}
\newtheorem{thm}{Theorem}

\newtheorem{lem}{Lemma}

\newtheorem{definition}{Definition}

\newcommand{\ba}{\begin{eqnarray}}
\newcommand{\ea}{\end{eqnarray}}
\newcommand{\bann}{\begin{eqnarray*}}
\newcommand{\eann}{\end{eqnarray*}}
\newcommand{\bal}{\begin{equation}\begin{aligned}}
\newcommand{\eal}{\end{aligned}\end{equation}}
\newcommand{\dm}[1]{\ketbra{#1}{#1}}

\newcolumntype{L}[1]{>{\raggedright}p{#1}}
\newcolumntype{C}[1]{>{\centering}p{#1}}
\newcolumntype{R}[1]{>{\raggedleft}p{#1}}
\newcolumntype{Y}{>{\centering\arraybackslash}X}

\usepackage{bm}

\newcommand{\nin}{{n_\text{\rm q}}}

\newcommand{\kin}{{k_\text{\rm q}}}
\newcommand{\din}{{d_\text{\rm q}}}
\newcommand{\eout}{{\varepsilon_{\rm out}}}
\newcommand{\etarg}{\varepsilon_{\rm targ}}
\newcommand{\ein}{{\varepsilon_{\rm in}}}

\newcommand{\al}{\alpha}

\newcommand{\gpred}{\gamma_\text{\rm pre}}

\newcommand{\eoutqag}{\varepsilon_{\rm out}^{\rm qAG}}

\newcommand{\codepar}[1]{\ensuremath{[\![#1]\!]}}

\newcommand{\eq}[1]{(\hyperref[eq:#1]{\ref*{eq:#1}})}

\renewcommand{\sec}[1]{\hyperref[sec:#1]{Section~\ref*{sec:#1}}}
\newcommand{\thrm}[1]{\hyperref[thrm:#1]{Theorem~\ref*{thrm:#1}}}
\newcommand{\lemm}[1]{\hyperref[lemm:#1]{Lemma~\ref*{lemm:#1}}}
\newcommand{\prop}[1]{\hyperref[prop:#1]{Proposition~\ref*{prop:#1}}}
\newcommand{\corr}[1]{\hyperref[corr:#1]{Corollary~\ref*{corr:#1}}}
\newcommand{\fig}[1]{\hyperref[fig:#1]{~\ref*{fig:#1}}}
\newcommand{\deff}[1]{\hyperref[deff:#1]{~\ref*{deff:#1}}}
\renewcommand{\appendix}{\par
  \setcounter{section}{0}
  \setcounter{subsection}{0}
  \gdef\thesection{\Alph{section}}
}

\usepackage{hyperref}
\hypersetup{
    pdfstartview={FitH},
    colorlinks=true, 
    linkcolor=Violet, 
    citecolor=OliveGreen, 
    urlcolor=Magenta 
}
\usepackage{cleveref}
\crefname{section}{Appendix}{Appendices}
\crefname{subsection}{Appendix}{Appendices}
\crefname{thm}{Theorem}{Theorems}
\crefname{lem}{Lemma}{Lemmas}
\crefname{figure}{FIG.}{FIGs.}
\Crefname{figure}{FIG.}{FIGs.}

\crefname{equation}{Eq.}{Eqs.}
\Crefname{equation}{Eq.}{Eqs.}
\begin{document}

\title{Asymptotic magic state distillation with almost linear rate}

\author{Koki Ehara}
\email{kokieha0331@g.ecc.u-tokyo.ac.jp}
\affiliation{Department of Basic Science, The University of Tokyo, 3-8-1 Komaba, Meguro-ku, Tokyo 153-8902, Japan}

\author{Ryuji Takagi}
\email{ryujitakagi@g.ecc.u-tokyo.ac.jp}
\affiliation{Department of Basic Science, The University of Tokyo, 3-8-1 Komaba, Meguro-ku, Tokyo 153-8902, Japan}

\begin{abstract}
The overhead exponent---characterizing the scaling of the number of noisy magic states with respect to the target distillation error---has been a central quantity to benchmark magic state distillation protocols. On the other hand, a related but less investigated quantity motivated by an information-theoretic viewpoint is the asymptotic distillation rate, the largest ratio of output to input magic states such that error vanishes asymptotically. These two quantities are tightly related in the specific case---the overhead exponent is zero if and only if the asymptotic distillation rate is linear. However, their relationship in other regimes has been unclear. Here, we show that their quantitative relation is generally not robust, by presenting a family of magic state distillation protocols with an overhead exponent not close to zero---in fact, larger than one---that still achieves the asymptotic rate arbitrarily close to the linear rate. This implies that the distillation rate is not constrained by the overhead exponent within the sublinear rate regime. Notably, our protocol is based on error checking by measurements of logical Clifford operators, which underlies the recent magic state cultivation protocol, suggesting the potential of this mechanism for asymptotic magic state distillation.
\end{abstract}

\maketitle

\paragraph{Introduction.---}

On evaluating the performance of the magic state distillation protocol ~\cite{Bravyi2004-jc,Goto2014-ja,Bravyi2004-jc,Itogawa2025-ul,Haah2017-gl,Haah2017-lj,Hastings2018-oi,Wills2024-im,Scruby2025-jw,Golowich2024-ic,Meier2012-ip,Nguyen2024-fi,Krishna2019-xs,Krishna2018-os,Litinski2019-bq,Prakash2020-ty,Reichardt2005-iu,Hastings2018-oi,Campbell2012-qx,Kalra2026-oq,Prakash2025-zd,Reichardt2005-iu,Anwar2012-kw,Krishna2018-os}, one needs to set the relevant figures of merit.  
When a fault-tolerant quantum computation (FTQC) algorithm requires magic states, both the required error precision and the number of states must be specified.  
In such situations, a performance indicator that evaluates the trade-off between the error suppression rate and the distillation yield becomes effective.  
One such indicator arises from the fact that, when using an $\codepar{n,k,d}$ code to prepare a magic state with error rate suppressed below $\varepsilon$, one typically needs $O(\log^{\gamma}(1/\varepsilon))$ input magic states, where $\gamma=\log_d (O(n)/k)$ is called the \emph{overhead exponent}.  
From this perspective of realizing FTQC with the lowest possible overhead, extensive research has been conducted on minimizing the value of $\gamma$ \cite{Hastings2018-oi,Bravyi2012-us,Krishna2019-xs,Wills2024-im,Golowich2024-ic,Nguyen2024-fi}.

However, the overhead exponent is not the only choice of performance quantifier.
Under the setting where one has access to a large number of noisy magic states and aims to maximize the yield of pure magic states, investigating the \emph{asymptotic distillation rate}---the maximum rate of output to input magic states under the condition that error vanishes in the asymptotic limit---is also motivated. 
In fact, this is the standard quantity studied in information theory, such as source and channel coding~\cite{wilde2013quantum}, as well as quantum resource theories in the context of resource distillation~\cite{RevModPhys.81.865}.
The difference from the overhead exponent is that the asymptotic distillation rate does not take into account the speed of error suppression but aims to maximize resource yield guaranteeing that error can be made as small as one wishes by taking a sufficiently large number of input copies.   

As both overhead exponent and distillation rate quantify the efficiency of magic state distillation, it is reasonable to expect some relation between these two quantities.
Indeed, one can observe that a magic state distillation protocol has zero overhead exponent if and only if the distillation rate becomes linear, i.e., from $n$ copies of noisy magic states, the protocol can produce $rn$ copies of purer magic states with $r>0$ in a way that the error vanishes in the $n\to\infty$ limit.
The existence of such a protocol has recently been discovered~\cite{Wills2024-im}, realizing the ``constant overhead'' magic state distillation with $\gamma=0$, and hence the linear magic state distillation rate.
However, the relationship between these two quantities has been unclear except for this specific case.
Given the above observation, it is natural to investigate their neighborhood---namely, whether the protocol with an almost linear rate always comes with an almost zero overhead exponent.
Besides the theoretical interest, this question is also motivated from the viewpoint of implementation, as error-correcting codes with $\gamma\sim 0$ typically entail a highly complex code design~\cite{Nguyen2024-fi,Scruby2025-jw,Krishna2018-os}.

Here, we show that the relationship between the overhead exponent and distillation rate dramatically changes for the regime of non-zero overhead exponent, or equivalently, sub-linear distillation rate. 
We achieve this by explicitly presenting a family of magic state distillation protocols that converts $n$ copies of noisy magic states to $a_n$ pure magic states with vanishing error for an arbitrary choice of $a_n=o\qty(\frac{n}{\log n})$---particularly including $O(n^{1-\eta})$ for an arbitrary $\eta>0$---while the overhead exponent $\gamma$ being $\gamma>1$.
This showcases the discontinuous jump of the overhead exponent despite the smooth transition in terms of the asymptotic distillation rate.

The protocol we employ is based on the generalized Hadamard-test protocols established in Ref.~\cite{Haah2017-mz}, and this brings another interesting insight. 
One can observe that magic state distillation protocols can typically be categorized into two approaches~\cite{Haah2017-mz}.
The \emph{transversal-type} protocol, originating from Ref.~\cite{Bravyi2004-jc}, distills the magic state by transversally implementing a non-Clifford gate~\cite{Bravyi2012-ui,Shi2024-kd,Golowich2025-rg,Nguyen2024-fi}. 
Since the discovery of quantum error-correcting codes that achieve sublogarithmic overhead by Hastings \textit{et al.}~\cite{Hastings2018-oi}, this type of protocol has been widely studied as part of the ongoing effort to find constructions with smaller values of $\gamma$, including the first protocol with sublogarithmic overhead~\cite{Hastings2018-oi} and the recent discovery with constant overhead~\cite{Wills2024-im}. 
On the other hand, the \emph{Hadamard-test-type} approach, proposed in Ref.~\cite{Knill2004-mz} and later generalized by Ref.~\cite{Haah2017-mz}, focuses on the fact that the magic state is an eigenstate of a Clifford gate and performs projective measurements to drive the system toward the magic state. 
Although $\gamma>1$~\cite{Haah2017-mz}, this approach has proven effective for the magic state purification in practical regimes, such as magic state cultivation \cite{Gidney2024-br,chen2025efficientmagicstatecultivation,Sahay2026-jy,Vaknin2026-hd,Hirano2025-yh,Claes2025-rm}.
Our result, which employs the latter approach, suggests that Hadamard-test-type protocols can maintain a favorable yield not only in practically relevant error-suppression regimes but also in regimes requiring even higher levels of error suppression.

To further support this view and see how our protocol performs in concrete settings, we also numerically compute the distillation rate to achieve the desired accuracy in the extensive error regime. As a result, we find that our protocol performs favorably in a wide range of target error rate compared to a construction based on the algebraic geometry code achieving constant rate~\cite{Wills2024-im}, as well as a protocol based on concatenation of the 15-qubit code~\cite{Bravyi2004-jc}.

 \emph{Setting}---
Magic state distillation is a process consisting of Clifford operations that transform noisy magic states to purer magic states. 
The reasoning behind this is that Clifford operations are typically implementable in a fault-tolerant manner in many standard error-correcting codes, and thus the error on Clifford operations can easily be suppressed to the level that is negligible compared to the error on non-Clifford gates.
This observation allows one to assume that Clifford operations can be implemented perfectly, and we follow this standard assumption in this work.
Specifically, a magic state distillation protocol is specified by a series $\{\Lambda_n\}_n$ of Clifford operations, and we say that the distillation protocol achieves the rates $\qty{\frac{a_n}{n}}_n$ with error $\{\varepsilon_n\}_n$ if 
\begin{align}
 \frac{1}{2}\norm{\Lambda_n(\rho^{\otimes n})-\dm{T}^{\otimes a_n}}_1\leq \varepsilon_n
\end{align}
where $\norm{\cdot}_1$ is the trace norm, $\rho$ is a noisy magic state as input states, and $\ket{T}\coloneqq \frac{1}{\sqrt{2}}(\ket{0} + e^{i\pi/4}\ket{1})$. 
(The detailed setup and notation are given in \Cref{app:def_of_msd}.)
The overhead exponent is the constant $\gamma$ that determines the scaling of the overhead $\frac{n}{a_n}$ of the noisy magic states with respect to the error $\varepsilon_n$ as 
$\frac{n}{a_n} = O\qty(\log^\gamma\qty(\frac{1}{\varepsilon_n}))$. 
Namely, the protocol with a small overhead exponent can reduce the error quickly by using a small number of noisy magic states relative to the output magic states.

On the other hand, if we have access to many copies of the noisy magic states, it is also motivated to maximize the rate under the condition that the error vanishes asymptotically. 
Formally, we say that the rates $\qty{\frac{a_n}{n}}_n$ are achievable if there exists a magic state distillation protocol achieving these rates with errors $\qty{\varepsilon_n}_n$ satisfying $\varepsilon_n\xrightarrow[n\to\infty]{}0$.  

These two performance quantifiers are closely related in the extreme case of $\gamma=0$. 
Indeed, when $\gamma=0$, the overhead $\frac{n}{a_n}$ becomes independent of $\varepsilon_n$ in the large $n$ limit, making it possible to keep $\frac{n}{a_n}$ constant while realizing $\varepsilon_n\xrightarrow[n\to\infty]{} 0$. 
This particularly means that the distillation rates $\qty{\frac{a_n}{n}}_n$ can be made $\frac{a_n}{n} \xrightarrow[n\to \infty]{} R>0$ with some constant $R$, i.e., realizing a positive linear asymptotic rate. 
Conversely, if the linear asymptotic rate is nonzero, it must be the case that $\gamma=0$ because any $\gamma>0$ results in the diverging overhead $\frac{n}{a_n}$ in the limit of $\varepsilon_n\xrightarrow[n\to\infty]{}0$.

Thus, only a partial relationship is observed between the overhead exponent and the asymptotic distillation rate. We therefore investigate how the asymptotic distillation rate varies in the region where the overhead exponent $\gamma > 0$. 
We find that even in the region where the overhead exponent is large—--specifically, when $\gamma > 1$—--it is possible to approach a linear rate arbitrarily well up to a logarithmic factor.
 We show this by employing the Hadamard-test type magic state distillation protocol as described in the following.

 \emph{Hadamard-test type distillation protocol}---
This type of protocol is based on the fact that the magic state---and Clifford equivalent thereof---is (+1)-eigenstate of a Clifford gate, such as the Hadamard gate. 
That is, by repeatedly performing the Hadamard test using a controlled-Hadamard gate, one can extract magic states with high fidelity. 
Moreover, since the controlled-Hadamard gate is not a Clifford gate, it is decomposed into the form of Clifford $+ T$, and the $T$ gate is implemented by the gate teleportation circuit, which consists of Clifford operations. Such protocols originate from the work of Knill, where a protocol based on the $\codepar{7,1,3}$ Steane code was proposed~\cite{Knill2004-mz}. It was later generalized to constructions using arbitrary weakly self-dual CSS codes~\cite{Haah2017-mz}. Since a weakly self-dual code satisfies $H_X = H_Z$ (where $H_X$ and $H_Z$ are parity-check matrix for $X$ and $Z$ stabilizer), the Hadamard gate can be implemented transversally. A circuit-level description of this construction and the relevant property of weakly self-dual CSS codes are summarized in \Cref{app:def_of_msd,appendix_addressability}.

One can consider repeatedly performing the Hadamard test in order to achieve $\din$-th order error reduction on a quantum error-correcting code with such parameters. In magic state distillation on a quantum error-correcting code with $\kin$ logical qubits and code distance $\din$, the error transforms as follows with each round of the Hadamard test.
\begin{align}\label{eq:error_reduction}
    \eout=O\qty(\ein^2)+O\qty(\ein^\din).
\end{align}
Here $\ein$ and $\eout$ are the input and output error rates. The first term in \Cref{eq:error_reduction} corresponds to the case where the two $T$ states constituting a single Hadamard test for a magic state simultaneously suffer from errors. (See \Cref{app:def_of_msd} for more details.)
To eliminate this error, we introduce additional redundancy in the execution of the Hadamard test.
We represent this redundancy using the $m$ by $a_n$ parity-check matrix $M$ of a classical error-correcting code. It is designed to detect errors in the measurements of magic states encoded in a quantum error-correcting code. Specifically, the matrix $M$ determines the participation of the $a_n$ noisy magic states across the $m$ rounds of Hadamard tests. If the $(j,i)$-th entry is $1$, then the $i$-th encoded magic state $\rho$ participates in the $j$-th measurement.  To guarantee $\din$-th order error suppression, it is sufficient to require that $|v|+2|Mv|\ge \din$ for every nonzero binary vector $v\in \qty{0,1}^{a_n}$ with $|v|\le \din -1$. Here, $|Mv|$ denotes the number of violated parity checks associated with the error pattern $v$. While the code distance property only guarantees $|Mv|\ge 1$, this is insufficient because parity-check measurements themselves may fail. In particular, hiding a violated check requires at least two additional faults from the measurement circuit implementation. Hence, if an error pattern $v$ violates $|Mv|$ checks, an undetected process requires at least $|v|+2|Mv|$ faults in total. Therefore, requiring $|Mv|\ge (\din-1)/2$ for all $1\le |v|\le \din-1$ is sufficient to ensure that any undetected process has weight at least $\din$.
For such a parity-check matrix, it is known that there exist parameters $m, a_n, \kin$ satisfying the following relation, under which the error of the encoded state can be suppressed to $O(\ein^{\din})$ after measurement with post-selection~\cite{Haah2017-mz}.
\begin{align}\label{eq:relationship_with_an_m}
    \frac{m}{\frac{\qty(\din-1)}{2}}=\frac{a_n}{\kin}.
\end{align}
The formal condition underlying this relation is given in \Cref{app:outer_code}.
Therefore, the overhead exponent of such a protocol based on the Hadamard test can be computed as follows.
\begin{align}
    \gamma=\frac{\log\frac{n}{a_n}}{\log \din}\label{eq:gamma_1}
    =\frac{\log\frac{a_n+2m\nin}{a_n}}{\log\din} = \frac{\log\qty(1+\frac{\qty(\din-1)\nin}{\kin})}{\log\din}
\end{align}
where the last equality is because of 
\Cref{eq:relationship_with_an_m}.

    Here, since $\codepar{\nin,\kin,\din}$ are the parameters of a quantum error-correcting code and hence $\nin>\kin$ and $\din>1$, the overhead exponent $\gamma$ in Hadamard-test-type magic state distillation can be evaluated as follows.
\begin{align}
    \gamma=\frac{\log\qty(1+\frac{\qty(\din-1)\nin}{\kin})}{\log\din}>\frac{\log\qty(1+\qty(\din-1))}{\log\din}=1
\end{align}
That is, existing Hadamard-test-type protocols satisfy $\gamma > 1$.

 \emph{Distillation with almost linear rate with $\gamma>1$}---
We now investigate the limits of the asymptotic distillation rate achievable using the Hadamard-test-type protocol described in the previous section. We first summarize our main results in the following \Cref{thm:asymptotic_sublinear}.
\begin{thm}\label{thm:asymptotic_sublinear}
For given arbitrary series $\qty{\frac{a_n}{n}}_n$ such that $\frac{a_n}{n} = o\qty(\frac{1}{\log n})$, there exists a magic state distillation protocol that achieves the rates $\qty{\frac{a_n}{n}}_n$ with vanishing error $\varepsilon_n\xrightarrow[n\to\infty]{} 0$, while having the overhead exponent $\gamma>1$. 
\end{thm}

We show this by analyzing an explicit magic state distillation protocol. 
The main property of the protocol needed to achieve our goal is the flexibility in controlling the tradeoff between the encoding rates and distance of the quantum error-correcting codes detecting the logical error in the noisy magic states, which would allow us to choose a high encoding rates while the error eventually vanishes in the large $n$ limit. 

\begin{figure}
    \centering
    \includegraphics[width=\linewidth]{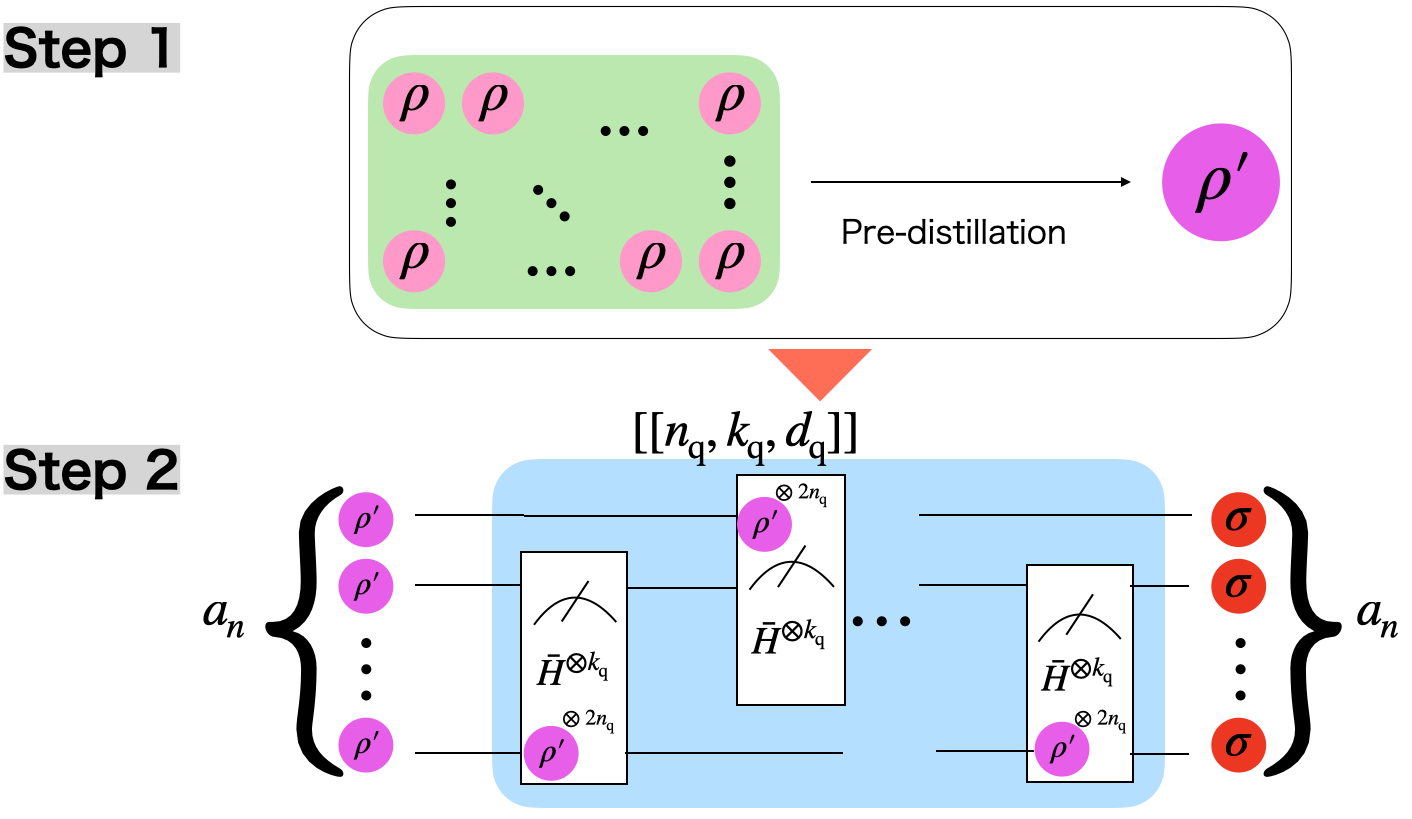}
    \caption{Overview of the distillation protocol. Step 1 is the pre-distillation step, which generates intermediate magic states from the initially prepared magic states so that they can withstand the distillation performed in Step 2.  
After that, the resulting states are processed in Step 2 to achieve $\eout \to 0$.
}
    \label{fig:distilaion_step}
\end{figure}
Our strategy is as follows: We divide our entire distillation protocol into two steps as shown in \Cref{fig:distilaion_step} and \Cref{eq:distill_step}.
\begin{align}\label{eq:distill_step}
    \rho^{\otimes n}\xrightarrow[\Lambda_{\mathrm{pre}}]{\ein\to\ein'}\rho'^{\otimes n'}\xrightarrow[\Lambda_{n'}]{\ein'\to\eout}\sigma^{\otimes a_{n}},
\end{align}
In the first step, we perform a process called \emph{pre-distillation}, which slightly reduces the noise in the initial magic states $\rho$ to produce intermediate magic states $\rho'$. This first step ensures that the error rate of the distilled magic states remains below the threshold of the quantum error-correcting code employed in Step 2. In this paper, we employ the $15 \to 1$ protocol based on the Reed-Muller code \cite{Bravyi2004-jc,Bravyi2012-ui} for this step. This choice is motivated by the protocol's high fault-tolerant threshold, which makes it particularly effective for the initial distillation of noisy magic states. 
Subsequently, we proceed to Step 2, where we execute magic state distillation to suppress the error to be vanishingly small. We consider a family of weakly self-dual CSS codes $\codepar{\nin,\kin(\nin),\din(\nin)}$ with odd distance and high encoding rate, where $\kin(\nin)/\nin \xrightarrow[\nin\to\infty]{}1$ and growing distance, i.e., $\din(\nin)\xrightarrow[\nin\to\infty]{}\infty$.
In particular, for weakly self-dual CSS codes with odd distance, the following lemma is known.

\begin{lem}[\cite{Haah2017-mz,Tansuwannont2025-kt}]\label{lem:non-addressability}
For an arbitrary weakly self-dual CSS code with odd distance, there exists a logical basis such that the logical Hadamard gate $\Bar{H}$ acts transversally as $H^{\otimes \nin}= \Bar{H}^{\otimes \kin}$.
\end{lem}

That is, by choosing a code with odd distance, all encoded logical states can participate in the Hadamard test with a single transversal gate. Encoding into an odd-distance weakly self-dual CSS code allows $\kin$ of $a_n$ noisy magic states to participate in a joint Hadamard test, where the parity check matrix $M$ governing the test schedule has a row Hamming weight of $\kin$. In the $j$-th Hadamard test, we refer to the $j$-th row of $M$ and encode $\kin$ noisy magic states, selected from $a_n$ inputs, into a quantum error-correcting code. After performing the Hadamard-test-based check, we apply the inverse encoding circuit. By repeating this process $m$ times (the number of rows in $M$), we achieve $\din$-th order error reduction.

By evaluating $\frac{a_n}{n}$ in the limit $\din \to \infty$ for such protocols, we obtain the desired relationship.

 In our approach, distillation is achieved by a single protocol based on scalable high-rate codes.
As an existence theorem for high-rate codes among such weakly self-dual CSS codes, the Gilbert–Varshamov bound \cite{Ashikhmin1999-xw,Calderbank1996-fh} is known.
Under this existence theorem, in order to obtain codes with a high encoding rate, we choose parameters such that $h(\din/\nin)\xrightarrow[\nin\to \infty]{}0$, where $\nin$ is the number of physical qubits and $h(x)\coloneqq-x\log_2(x)-(1-x)\log_2(1-x)$ is the binary entropy function. 
For ease of analysis, we introduce an auxiliary function $\al(\nin)$ and write $\din = \al(\nin)\nin$. In this case, in order for the code parameters to satisfy the required conditions, we impose $\al(\nin)\xrightarrow[\nin\to\infty]{}0$ and $\al(\nin)\nin\xrightarrow[\nin\to\infty]{}\infty$.
Based on this, the code used for magic state distillation in our approach can be written as follows.
\begin{align}
    \left\llbracket\nin, \kin=\nin\qty(1-h\qty(\al\qty(\nin))),\din=\al\qty(\nin)\nin=\omega(1)\right\rrbracket.
\end{align}

As described above, the test schedule of the Hadamard test is specified by the parity-check matrix $M$ satisfying \Cref{eq:relationship_with_an_m}.
The existence of such a classical parity-check matrix as in \Cref{eq:relationship_with_an_m} is guaranteed by the existence of a Tanner graph in which the number of bit nodes is $a_n$, the number of checks is $m$, and the respective degrees are $(\din - 1)/2$ and $\kin$. In this case, combinatorial studies \cite{Furedi1995-iv} have shown that the scaling of $a_n$ for which such a biregular bipartite graph exists is given by $a_n =O\qty(\mathrm{poly}\qty(\kin^{\din}))$. 
The fact that the growth of $a_n$ exceeds a certain rate becomes important in the subsequent analysis of the asymptotic rate, on which we elaborate in \Cref{app:Efficiency of Hadamard type magic state distillation}.

We also choose the intermediate error rate $\ein'$ after pre-distillation to vanish in the asymptotic limit, because the number of possible fault patterns increases with the sizes of the inner and outer codes.  The reason this does not spoil the rate is that the cost of preparing such an intermediate state grows only polylogarithmically in $1/\ein'$, and hence contributes only additional distance-dependent factors in the denominator of the rate estimate.

Having outlined the overall procedure, we now briefly describe the underlying logic for analyzing the magic state resource yield before and after the distillation process. For a derivation and detailed calculations, see \Cref{app:proof of thm}. First, the cost $n=C_p n'$ required for magic state distillation under this scheme, where $C_p$ is the cost of pre-distillation, can be characterized as follows 
\begin{align}
    n' &=  a_{n} + 2\nin \cdot m,\label{eq:cost_of_main}\\
    C_p&=C_p'\qty(\log^{2.46}\qty(\frac{1}{\ein'})),\label{eq:cost_of-predistill}
\end{align}
where,  $C_p'$ is a constant and \Cref{eq:cost_of-predistill} follows from the fact that distillation based on the Reed–Muller code has $\gamma = 2.46$.
 Then
\begin{align}\label{eq:distillation-rate_ours}
    \frac{a_{n}}{n}= \frac{a_n}{ C_p'\qty(\log^{2.46}\qty(\frac{1}{\ein'}))\qty(a_{n} + 2\nin \cdot m)}.
\end{align}
In order to study the asymptotic regime, we evaluate the following quantity:
\begin{align}\label{rate_limit}
\min\qty{t: \lim_{n\to\infty} \frac{a_n^t}{C_p \qty(a_{n} + 2\nin \cdot m)} > 0 }.   
\end{align}
In \Cref{rate_limit}, the parameter $t$ characterizes the optimal scaling behavior of the rate in the asymptotic regime.

Substituting \Cref{eq:cost_of-predistill} into \Cref{rate_limit}, one can choose the parameters so that $t\to1$, implying that the rate can be made arbitrarily close to linear.

Finally, since the magic state distillation protocol considered in this paper employs post-selection, the effective rate $R^{\mathrm{eff}}_{n}$ is given by the actual rate multiplied by the success probability.
\begin{align}
    R^{\mathrm{eff}}_{n}=(1-\ein')^{n'}\frac{a_{n}}{n}.
\end{align}
In \Cref{app:proof of thm}, we show that the success probability $(1-\ein')^{n'}$ converges to 1 in the limit $n \to \infty$. Therefore, since the success probability converges to $1$, the effect of post-selection on the asymptotic distillation rate becomes negligible. Note that the cost arising from the success probability of post-selection in pre-distillation is included in $C_p$.

\paragraph{Distillation rate in finite size distillation---}
Here, we investigate how our protocol performs in the non-asymptotic regime. 
Detailed descriptions of the setting can be found in \Cref{app:finite_size}.

\begin{figure}[h]
    \centering
    \includegraphics[width=1\linewidth]{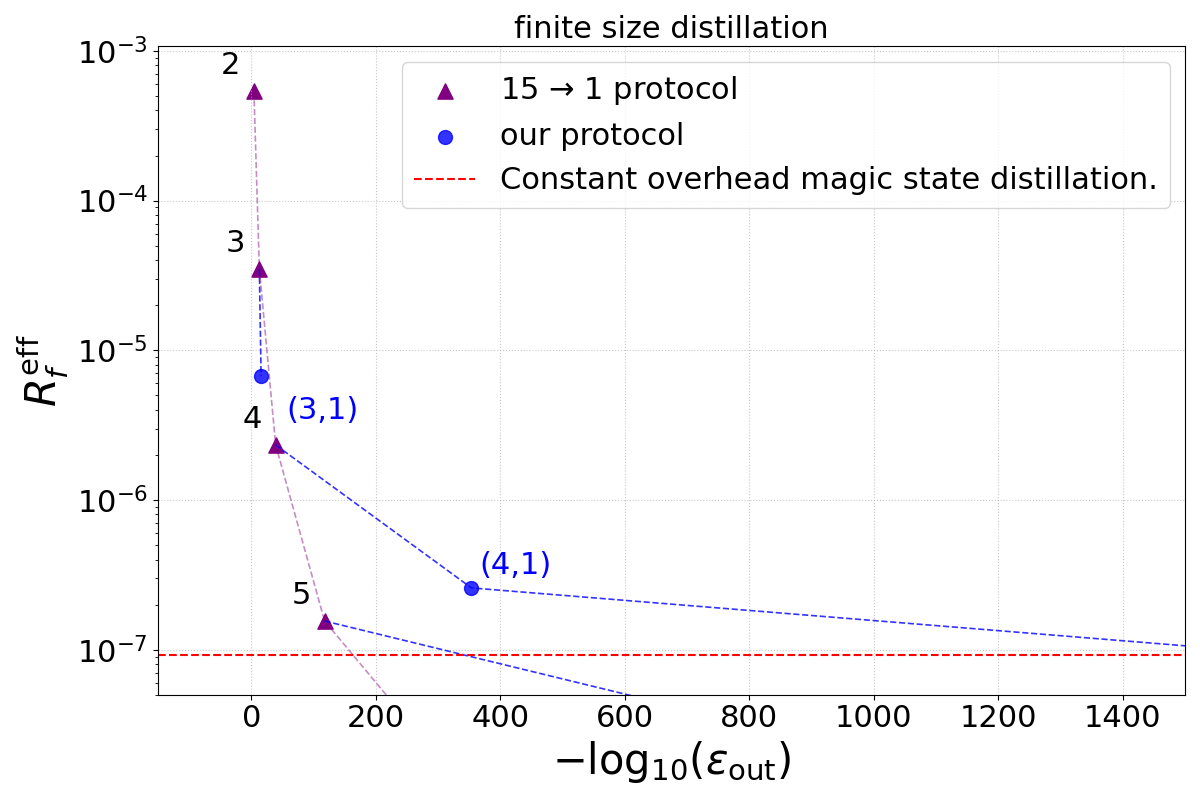}
    \caption{Numerical plots for finite-size magic state distillation. The vertical axis shows the distillation rate, while the horizontal axis shows the exponent of the output error. The purple triangles represent repeated applications of the $15\to1$ pre-distillation protocol, and the blue circles are obtained by subsequently applying our protocol. A blue label $(p,r)$ indicates that $r$ rounds of our protocol are performed after $p$ rounds of $15\to1$ pre-distillation. The red dashed line indicates the rate of magic state distillation with constant overhead. For detailed information such as the coordinates of the points in this figure, see \Cref{app:finite_size}.
}
    \label{fig:comparison_with_quantum algebraic geometry}
\end{figure}

Consider distilling a magic state $\ket{T}$ with an initial error rate of $0.1$ into an output error rate $\eout$. 
In \Cref{fig:comparison_with_quantum algebraic geometry}, we plot the output logical error rate and distillation rate of our protocol that uses $15\to 1$ protocol with Reed-Muller code for pre-distillation. 

We label each concatenated protocol by a pair $(p,r)$, where $p$ is the number of times the $15\to1$ pre-distillation protocol is applied and $r$ is the number of times our protocol is applied afterward.

For comparison, we also include in \Cref{fig:comparison_with_quantum algebraic geometry} the performance of the protocol in \cite{Wills2024-im}. 
Here, we focus on the smallest instance of the explicit family presented in \cite{Wills2024-im} for benchmarking, although it may not be the optimal instance among all possible protocols in their framework. 
As can be seen in \Cref{fig:comparison_with_quantum algebraic geometry}, our protocol achieves higher rates for a wide range of output error rates, all the way down to the extremely small value of $\sim10^{-350}$.

To see the region closer to the practical setting, we also compare with the $15\to 1$ protocol. 

It shows that our method becomes advantageous around $\eout \sim 10^{-19}$. From this point, it is in principle possible to search for more efficient protocols and aim for significantly smaller infidelity while maintaining the rate.  

Our finite-size analysis suggests that sublinear-rate protocols may occupy a distinct finite-size regime that is not well represented by the two benchmark instances considered here, namely repeated $15\to1$ distillation and a representative constant-overhead construction based on quantum algebraic geometry codes.
Nevertheless, our results exhibit particularly strong performance in the regime below an error rate of $10^{-40}$, and are therefore primarily theoretical in nature. However, the blue point $(3,1)$ in \Cref{fig:comparison_with_quantum algebraic geometry} corresponds to an error rate of approximately $10^{-19}$, which may still reasonably be regarded as a practical regime (see \Cref{app_fig:zoom_fig2} for a more detailed view of this region). This point demonstrates that it is a higher distillation rate than that achieved by executing the $15\to1$ protocol four times.

\paragraph{Conclusion.---}
We have shown, by explicitly constructing a protocol, that magic state distillation protocols with a large overhead exponent can still achieve a rate close to linear from the perspective of the asymptotic distillation rate. This implies a nontrivial separation between the overhead exponent and the asymptotic distillation rate. It suggests that, in the study of asymptotic transformations in the quantum resource theory of magic, one should not restrict their attention to protocols with small overhead exponents.

The protocol used to prove our main result is based on the Hadamard test, and may potentially be combined with magic state cultivation~\cite{Gidney2024-br} to further improve resource efficiency toward low-overhead FTQC. Indeed, our numerical comparison of finite-size distillation rates indicates that the proposed protocol performs most favorably in fidelity regimes that are inaccessible to magic state cultivation.

From a quantum information theoretic perspective, our results motivate the search for operational quantities that more finely characterize magic distillation in the sublinear rate regime, where the overhead exponent alone does not fully capture the achievable yield. 
Another important research direction is to clarify whether the nearly linear sublinear rate demonstrated here can be realized using quantum low-density parity-check codes as the underlying quantum codes. Moreover, even for finite-size distillation protocols, it remains unclear which combination of protocol parameters yields the optimal concatenated construction.

\paragraph{Acknowledgment.---}
This work was supported by JST SPRING, Grant Number JPMJSP2108, JST CREST Grant Number JPMJCR23I3, JST NEXUS Grant Number JPMJNX26C2, JSPS KAKENHI Grant Number JP24K16975, JP25K00924, JP26H02015, 26KJ0965.

\let\oldaddcontentsline\addcontentsline 
\renewcommand{\addcontentsline}[3]{}
\bibliographystyle{apsrmp4-2}
\bibliography{sublinear_arxiv4}

\let\addcontentsline\oldaddcontentsline

\clearpage
\newgeometry{hmargin=1.2in,vmargin=0.8in}

\widetext

\appendix
\renewcommand{\thesubsection}{\arabic{subsection}}
\setcounter{thm}{0}
\renewcommand{\thethm}{S.\arabic{thm}}
\setcounter{figure}{0}
\renewcommand{\thefigure}{S.\arabic{figure}}

\begin{center}
{\large \bf Appendices}
\end{center}

\tableofcontents

\section{Generalized Hadamard-test type magic state distillation}
\subsection{Setup}\label{app:def_of_msd}
\begin{figure}[h]
  \begin{minipage}[b]{0.33\linewidth}
   \centering
        \includegraphics[keepaspectratio, scale=0.8]{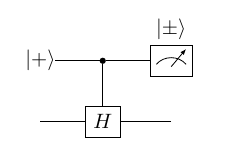}
    \caption{Stabilizer measurement for eigenstate of Hadamard gate.}\label{app_fig:CH_check}

  \end{minipage}
  \begin{minipage}[b]{0.33\linewidth}
          \begin{flushleft}
                   \includegraphics[keepaspectratio, scale=0.75]{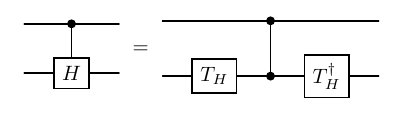}
          \caption{Decomposition of Contorol-Hadamard gate.}
          \label{app_fig:decomposition_of_CH}
          \end{flushleft}

  \end{minipage}
  \begin{minipage}[b]{0.33\linewidth}
    \begin{flushright}
        \includegraphics[keepaspectratio, scale=0.75]{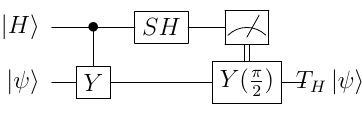}
    \caption{Gate teleportation of $T_H$ gate}\label{app_fig:injection_CH}
    \end{flushright}
    
  \end{minipage}
\end{figure}

In this section, we explain magic state distillation introduced in \cite{Haah2017-mz} based on the Hadamard test and discuss its distillation efficiency.
The definition and notation of this paper are as follows. 
\begin{align}
X = \begin{pmatrix} 0 & 1 \\ 1 & 0 \end{pmatrix}, \quad
Y = \begin{pmatrix} 0 & -i \\ i & 0 \end{pmatrix}, \quad
Z = \begin{pmatrix} 1 & 0 \\ 0 & -1 \end{pmatrix},\\
H = \frac{1}{\sqrt{2}}\begin{pmatrix} 1 & 1 \\ 1 & -1 \end{pmatrix}, 
\quad
T_H = e^{-i \pi Y/8} = 
\begin{pmatrix} 
\cos \frac \pi 8 & - \sin \frac \pi 8 \\
\sin \frac \pi 8 &   \cos \frac \pi 8 
\end{pmatrix},
\quad
T =
\begin{pmatrix} 
1 & 0 \\
0 &  e^{\frac{i\pi}{4}}
\end{pmatrix}.
\end{align}
We describe the methods of magic state distillation we used in this paper. In this paper, we use the Hadamard test based magic state distillation protocol. It is based on the fact that the magic state is $(+1)$ -eigenvalue of the Clifford gate. Especially,  $\ket{H}$ is $(+1)$ -eigenvalue of the Hadamard gate:
\begin{align}
     \ket{H}&=\cos\qty(\frac{\pi}{8})\ket{0}+\sin\qty(\frac{\pi}{8})\ket{1},\\
     H\ket{H}&=\ket{H}.
\end{align}
Moreover, the error model used in this paper is the stochastic error model. That is, it means that a $Y$ error occurs probabilistically. This is because any quantum channel can be projected onto Pauli noise via a twirling operation over the Clifford group.
Thus we can write $\rho$ as
\begin{equation}\label{app_eq:noisy_magic_state}
    \rho= (1-\ein)\ketbra{H}{H}+\ein \ketbra{-H}{-H},
\end{equation}
where $\ein$ is input error rate and $\ket{-H}$ is the $(-1)$ eigenstate of operator $H$.
We implement stabilizer measurement as \Cref{app_fig:CH_check}. Magic state distillation must be constructed solely from magic states and stabilizer operations.
Throughout this work, stabilizer operations refer to operations generated by Clifford unitaries, Pauli measurements, preparation of stabilizer states, and classical feedforward. 
Consequently, non-Clifford gates such as the $CH$ gate must be decomposed into gate teleportation as \Cref{app_fig:decomposition_of_CH} using magic states together with stabilizer operations.

\begin{figure}[h]
    \centering
    \includegraphics[width=0.75\linewidth]{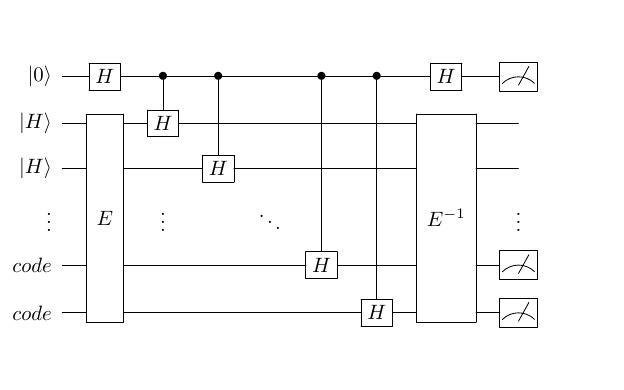}
    \caption{Quantum circuit for performing the Hadamard test of the $H$ gate on a quantum error-correcting code. The box $E$ represents the encoding circuit of the quantum error-correcting code, and $E^{-1}$ denotes its inverse circuit. The wires labeled $code$ represent the redundant qubits of the quantum error-correcting code used to perform distillation, and correspond to $\ket{0}$.
}
    \label{app_fig:Hadamard_test_on_QECC}
\end{figure}
Noisy magic states $\ket{T_H}$ are encoded into a quantum error-correcting code in the manner shown in \Cref{app_fig:Hadamard_test_on_QECC}, and the Hadamard test using the $H$ gate is performed via a transversal $H$ gate. Afterward, post-selection based on the measurement outcomes of the ancilla system is carried out, completing the distillation step.
When the quantum error-correcting code used has parameters $\codepar{\nin,\kin,\din}$, up to $\qty(\din - 1)$ errors within the code can be detected. Therefore, after post-selection, the error of the magic state is expected to scale as $O(\ein^{\din})$, where $\ein$ denotes the input error rate. However, within this scheme, the construction of the $CH$ gate requires two magic states per physical qubit, as shown in \Cref{app_fig:decomposition_of_CH}. In this situation, when simultaneous errors occur on the two $T_H$ gates applied to the same physical qubit, the ancillary qubit is affected by an undetectable error in the quantum error-correcting code. This is explained by the following equation.
\begin{equation}\label{Eq:measurement_error}
    \ketbra{0}{0}\otimes I + \ketbra{1}{1}\otimes T_HYZ(T_HY)^{\dagger}
               = (Z\otimes I)(\ketbra{0}{0}\otimes I + \ketbra{1}\otimes T_HZT_H^{\dagger})
\end{equation}

The probability of this error pattern is $O(\ein^2)$ as a function of the input error rate $\ein$.
Because of \Cref{Eq:measurement_error}, the error rate of the entire system can be expressed as 
\begin{align}
    \eout =O(\ein^2)+O(\ein^d).
\end{align} Escaping from this limitation and achieving higher-dimensional error suppression requires repeated measurements. Consider a classical code that prescribes how measurements for $H$ operator are to be repeated across all encoded magic states. We refer to this code as the \emph{outer code}. To distinguish it from this, we refer to the quantum error-correcting code as the \emph{inner code}.
In the following subsections, we review the properties of the inner and outer codes required for the entire system to achieve $O(\ein^{\din})$.
\subsection{Property of inner code}\label{appendix_addressability}
This section is devoted to the analysis of logical gates associated with the quantum error-correcting code called inner code.
\begin{definition}[Inner code]\label{def:innercode}
    The inner code is a quantum error-correcting code used to detect errors in the magic states, and is denoted by $\codepar{\nin,\kin,\din}$ in the family of weakly self-dual CSS codes.
\end{definition}
By implementing the circuits represented in \Cref{app_fig:CH_check}--\Cref{app_fig:injection_CH} on a quantum error-correcting code, detecting errors, and performing post-selection, the errors in the encoded magic states can be reduced. Weakly self-dual CSS codes are known as quantum error-correcting codes that allow such circuits to be implemented transversally.
\begin{definition}[Weakly self-dual CSS code]\label{app_def:weakly_selfdual_CSS_code}
    We say that a CSS code is \emph{weakly self-dual} if it is defined by the same binary parity-check matrix for both $X$- and $Z$-type stabilizers:
\begin{align}\label{eq_app:codespace}
    \Pi_X = \Pi_Z = \Pi,
\end{align}
where $\Pi$ satisfies the self-orthogonality condition
\begin{align}
    \Pi \Pi^{T} = 0 \pmod 2.
\end{align}
Equivalently, the classical code $C = \mathrm{row}(H)$ is self-orthogonal, that is,
\begin{align}
    C \subseteq C^\perp.
\end{align}
\end{definition} 
Combining \Cref{eq_app:codespace} with the fact that conjugation by the operator $H$ exchanges the operators $X$ and $Z$, we find that the transversal Hadamard gate is an operator that preserves the code space on a weakly self-dual CSS code.
In the magic state distillation protocols considered in this paper, we focus on protocols that output many magic states. That is, many logical qubits are encoded within a single inner code. The existence of such high-encoding-rate weakly self-dual CSS codes is guaranteed by the following \Cref{app_lem:GVbound}.
\begin{lem}[Gilbert-Varshamov bound~\cite{Ashikhmin1999-xw,Calderbank1996-fh}]\label{app_lem:GVbound}
There exists a weakly self-dual CSS code characterized by the following parameters.
    \begin{align}\label{eq:GV_bound}
        \left\llbracket \nin,\kin= \nin\qty(1- h\qty(\frac{\din}{\nin})), \din\right\rrbracket.
    \end{align}
    Here $\nin$ is a number of physical qubit, $h(x)\coloneqq-x\log_2(x)-(1-x)\log_2(1-x)$ is the binary entropy function.
\end{lem}
That is, it suffices to employ as the inner code a weakly self-dual CSS code with odd distance whose existence is guaranteed by \Cref{app_lem:GVbound}.
In this context, a relevant question is whether addressability, namely the ability to perform individual operations on each logical qubit, is required as a property of the inner code for the protocol.
Although addressability is a convenient property, it is not required for proving \Cref{thm:asymptotic_sublinear} in our work.
In this paper, we require only the transversal property $H^{\otimes \nin}= \Bar{H}^{\otimes \kin}$ for the inner code as a logical gate property. This is because it suffices to encode $\kin$ magic states simultaneously and have these $\kin$ logical magic states participate in the Hadamard test simultaneously using a circuit such as that in \Cref{app_fig:Hadamard_test_on_QECC}.
It is known that weakly self-dual CSS codes with odd distance possess the above transversal property \cite{Haah2017-mz,Tansuwannont2025-kt}.

\subsection{Property of outer code}\label{app:outer_code}
In this section, we explain what kinds of outer codes can be constructed for a given inner code.
\begin{definition}[Outer code]\label{def:outercode}
The outer code serves to detect errors that have escaped into the ancilla system.
    The outer code is a classical code with a binary parity-check matrix, $M$, of size $m$ by $a_n$, which specifies which qubit will undergo a parity check. Here, $m$ is the number of parity checks performed.
    \begin{align}\label{eq:outercode}
   \left.
M=
\begin{array}{c}
\\
\\
m\\
\\
\\[-1ex]
\end{array}
\right\{
\quad
\underbrace{\left(
\begin{array}{ccccccc}
\cdots & 0 & \cdots & \overbrace{\boxed{1}}^{i-\text{th magic state}} & \cdots & 0 & \cdots \\
\vdots & \vdots & \ddots & \vdots & \ddots & \vdots & \vdots \\
\cdots & 0 & \cdots & 0 & \cdots & 1 & \cdots \\
\end{array}
\right)}_{a_{n}}
\quad
\begin{array}{l}
\end{array}. 
\end{align}
\end{definition}
The role of the outer code is to detect measurement-type errors that are not detected by the inner quantum code. 
As discussed around \Cref{Eq:measurement_error}, such an error arises from two faults in the implementation of a Hadamard-test check and gives an effective error on the check outcome. 
To suppress these contributions to order $O(\ein^{\din})$, we need a stronger property than the usual distance of a classical code, which we formulate in terms of sensitivity.

\begin{definition}[sensitivity]
An $m$-by-$a_n$ binary parity-check matrix $M$ is said to be $(\tilde d,s)$-sensitive if, for every nonzero bit vector $v\in\{0,1\}^{a_n}$ with $|v|\leq \tilde d$, one has
\begin{align}
    |Mv|\geq s .
\end{align}
In other words, every error pattern of weight at most $\tilde d$ violates at least $s$ parity checks.

\end{definition}
For $\din$-th error reduction, we consider the outer code to be
\begin{align}\label{eq;app_sensitivity}
    \qty(\tilde d,s)
    =
    \qty(\din-1,\left\lfloor\frac{\din-1}{2}\right\rfloor)
\end{align}
sensitive. 
Indeed, consider an error pattern $v$ of weight $|v|\leq \din-1$ on the $a_n$ output magic states. 
If this pattern violates $|Mv|$ checks, then hiding these violated checks requires at least two additional faults for each violated check. 
Thus an undetected process associated with $v$ requires at least $|v|+2|Mv|$ faults. 
The above sensitivity condition ensures that this number is at least $\din$, and hence such undetected errors contribute only at order $O(\ein^{\din})$.

\begin{figure}[H]
    \centering
    \includegraphics[width=\linewidth]{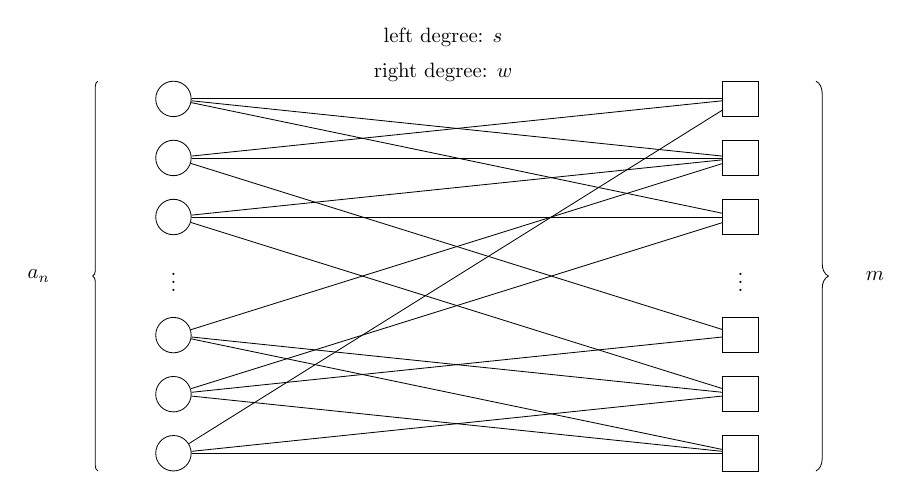}
   \caption{Tanner graph representation of the outer code. 
The circles on the left represent the $a_n$ magic states to be distilled, and the squares on the right represent the $m$ Hadamard-test checks. 
An edge indicates that the corresponding magic state participates in the corresponding check.}
    \label{fig:outercode_tannergraph}
\end{figure}

We now describe a convenient graph representation of such outer codes. 
The parity-check matrix $M$ can be represented by a Tanner graph whose left vertices correspond to the $a_n$ magic states and whose right vertices correspond to the $m$ Hadamard-test checks, as illustrated in \Cref{fig:outercode_tannergraph}. 
The degree of each check node is denoted by $w$, and the degree of each bit node is denoted by $s$. 
For a $(w,s)$-biregular Tanner graph, the number of edges can be counted in two ways, giving
\begin{align}\label{eq:outer_edge_count}
    \frac{m}{s}=\frac{a_n}{w}.
\end{align}
In our construction, we take the check weight to be $w=\kin$, since each Hadamard-test check acts on $\kin$ logical magic states encoded into one inner code block.

To guarantee the above sensitivity property, we choose the Tanner graph to have girth $g\geq 2(\din-1).$
If the girth were too small, low-weight error patterns could violate too few checks, and the required sensitivity would not be guaranteed. 
The existence of such biregular bipartite graphs is guaranteed by \cite{Furedi1995-iv}. 
In particular, for a family of inner codes with parameters 
$\codepar{\nin,\kin(\nin),\din(\nin)}$, one can choose the outer code size as
\begin{align}\label{eq:app_a_n}
    a_n=A(\nin)\kin(\nin),
\end{align}
where $A(\nin)$ grows polynomially in 
$\kin(\nin)^{\din(\nin)}$.
This scaling is used in the input-output count and asymptotic rate analysis below.

\section{Input-output count of Hadamard-test type magic state distillation}\label{app:Efficiency of Hadamard type magic state distillation}
\subsection{Overhead exponent of Hadamard-test type magic state distillation}
\begin{figure}[h]
    \centering
    \includegraphics[width=1\linewidth]{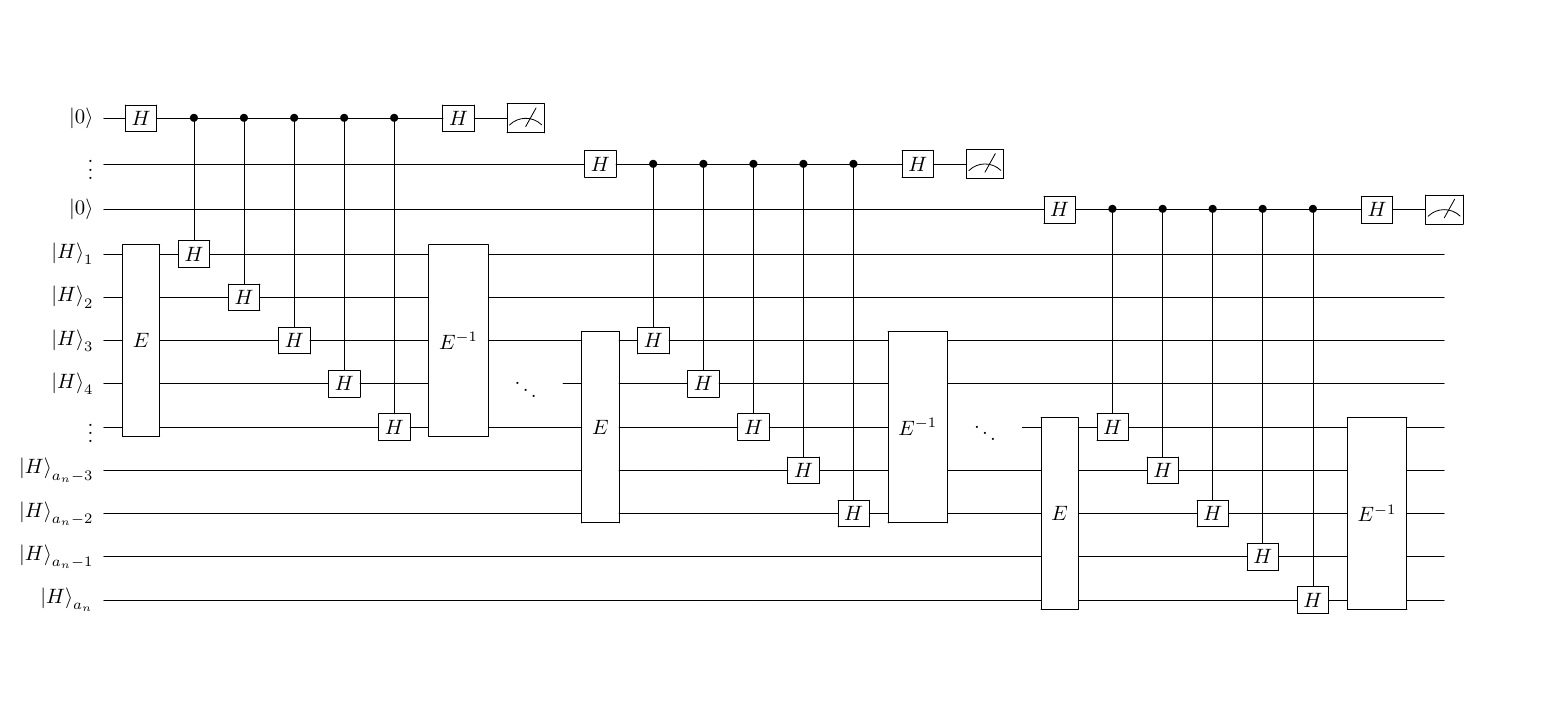}
    \caption{Circuit diagram of a protocol that achieves $O(\ein^{\din})$ by encoding noisy magic states into an inner code $\codepar{\nin,\kin,\din}$ according to the outer code and performing the Hadamard test. The wires corresponding to ancillas used for the redundancy of the inner code are omitted.
}
    \label{app_fig:allprotocol}
\end{figure}

In this section, we derive the input-output count and the overhead exponent of the Hadamard-test-type protocol when $\din$th-order error suppression is achieved by combining the inner code and outer code described above. First, we compute the input count based on \Cref{app_fig:allprotocol} and the discussion of the inner code and outer code in the preceding sections.

According to \Cref{app_fig:allprotocol}, the number of input magic states consumed in this Hadamard-test-type step is
\begin{align}\label{eq:app number of input magic states}
    n' = a_n + 2\nin m ,
\end{align}
where the first term corresponds to the $a_n$ noisy magic states to be distilled, and the second term comes from the two magic states required for each physical qubit in each Hadamard test.

We choose the outer code so that the $O(\ein^2)$ contribution is suppressed to $O(\ein^\din)$. 
For this purpose, we take the sensitivity to be as \Cref{eq;app_sensitivity}.
Moreover, since each Hadamard test acts on $\kin$ logical magic states encoded into the inner code, we set the row weight of the outer code parity-check matrix to be
\begin{align}\label{eq:app_weight_outercode}
    w=\kin .
\end{align}

We then obtain
\begin{align}
    n'
    &= a_n + 2\nin m \\
    &= a_n + 2\nin a_n\qty(\frac{s}{w}) \\
    &= a_n + a_n\qty(\din-1)\frac{\nin}{\kin} \\
    &= a_n\qty(\din + \qty(\din-1)\qty(\frac{\nin}{\kin}-1)).\label{eq:app_count_n'}
\end{align}
where the second line is due to \Cref{eq:outer_edge_count}, and in the third line we chose $s$ and $w$ as in \Cref{eq;app_sensitivity,eq:app_weight_outercode}.
We can now choose an appropriate inner code that meets our needs. 
We particularly choose a high-rate code such that $\frac{\kin}{\nin}=O(1)$ with odd distance, whose existence is guaranteed by the Gilbert–Varshamov bound for a weakly self-dual CSS code (\Cref{eq:GV_bound}).
This gives
\begin{align}\label{eq:app_inputcount}
    n' &=\din (1+o(1))a_n.
\end{align}
Up to this point, we have determined the number of magic states required to obtain $a_n$ magic states achieving $O(\ein^{\din})$. To conclude this section, we compute the overhead exponent of this protocol.

\begin{definition}[Overhead exponent]\label{app_def:overhead_exp}
    Let $n$ denote the number of input magic states, $a_n$ the number of output magic states, $d$ the distance of the quantum error-correcting code used for magic state distillation, and $\etarg$ the target output error rate. Then, the overhead required for distillation is $O(\log^{\gamma}(1/\etarg))$, here $\gamma$ can be described as follows.
    \begin{align}\label{eq:overhead_exp}
        \gamma=\frac{\log\qty(\frac{n'}{a_n})}{\log d}.
    \end{align}
    
\end{definition} 
According to this metric, the overhead exponent used in this paper is computed as follows from \Cref{eq:app_inputcount} and \Cref{eq:overhead_exp}.
\begin{align}\label{app_eq:overhead}
    \gamma=\frac{\log\frac{n'}{a_n}}{\log d}= \frac{\log\qty(\frac{\din(1+o(1))a_n}{a_n})}{\log \din}>\frac{\log \din}{\log \din}=1
\end{align}
That is, the magic state distillation protocol based on the Hadamard test has $\gamma > 1$.
Note that \Cref{app_eq:overhead} represents the overhead of the distillation protocol implemented by the combination of the inner code based on weakly self-dual CSS codes and corresponding outer codes, excluding pre-distillation.

\subsection{Analysis of sublinear rate protocol \label{app:proof of thm}} 
In this section, we analyze the achievable distillation rate by our protocol with $\gamma > 1$ introduced in the previous section.

 Our protocol consists of two parts: the pre-distillation part based on the $15\to1$ protocol, and the part that achieves $\eout\to0$ using a large inner code and outer code. It is therefore necessary to analyze the cost associated with each part. We first analyze the size of the required pre-distillation for a given inner code used in the distillation process.
    We prepare an inner code $\codepar{\nin, \kin = \nin(1 - 2h(\alpha(\nin))), \din=\alpha(\nin)\nin=\omega(1)}$. Here, the auxiliary function $\al(\nin)$ is a decreasing function satisfying $\al(\nin)\nin\xrightarrow[\nin\to\infty]{}\infty$.
The existence of such a weakly self-dual code is guaranteed by the Gilbert–Varshamov bound.  
For this inner code, there exists an outer code with an $m \times a_n$ parity-check matrix that achieves $\din$th-order error reduction, where the number of input magic states satisfies \Cref{eq:app_inputcount}.
In this case, the outer code has sensitivity $s = (\din-1)/2$, and $a_n$ satisfies $a_n = A(\nin)\kin$, where it is known that $A(\nin)$ is at most bounded by $A(\nin) < C_1 \kin^{C_2(\din-1)}$, where $C_1$ and $C_2$ are constant numbers.
It is also known that there exists a pair $(m, a_n)$ such that $m = a_n \cdot s / w$. Using these parameters, we analyze the distillation rate and obtain the desired theorem.

First, we determine a sufficient scaling of the intermediate input error rate $\ein'$ to the main Hadamard-test-type step. 
Here, $n'$ denotes the number of magic states consumed in this main step, before taking the pre-distillation cost into account. 
From \Cref{eq:outer_edge_count,eq:app number of input magic states}, we have
\begin{align}
    n'
    &= a_n+2\nin m \notag\\
    &= a_n+2\nin a_n\frac{s}{w}.
\end{align}
Using $s=(\din-1)/2$ and $w=\kin$, this becomes
\begin{align}
    n'
    &=a_n+a_n(\din-1)\frac{\nin}{\kin}.
\end{align}
For the high-rate inner code family used here, $\nin/\kin=1+o(1)$. 
Writing $a_n=A(\nin)\kin(\nin)$, we obtain
\begin{align}
    n'
    =
    A(\nin)\nin\din(1+o(1)).
    \label{app_eq:nprime_asymptotic}
\end{align}
Equivalently, since $\din=\alpha(\nin)\nin$, this scaling can be written as
\begin{align}
    n'
    =
    A(\nin)\alpha(\nin)\nin^2(1+o(1)).
    \label{app_eq:nprime_alpha}
\end{align}

To guarantee $\varepsilon_{\rm out}\to0$, it is sufficient to choose $\ein'$ so that the total probability of undetected fault patterns of weight at least $\din$ vanishes. 
Since all fault patterns of weight smaller than $\din$ are detected by the combination of inner code and outer code, the leading contribution starts with the order $(\ein')^\din$. 
We therefore consider the sufficient condition
\begin{align}
    \binom{n'}{\din}(\ein')^{\din}\xrightarrow[\nin\to\infty]{}0 .
    \label{app_eq:vanish_condition}
\end{align}
We now estimate the binomial coefficient in \Cref{app_eq:vanish_condition}. 
Substituting \Cref{app_eq:nprime_alpha} and $\din=\alpha(\nin)\nin$, and keeping only the leading exponential scaling by Stirling's approximation, we get
\begin{align}
    \begin{split}
        \binom{n'}{\din}
        &=
        \binom{A(\nin)\alpha(\nin)\nin^2(1+o(1))}
              {\alpha(\nin)\nin} \\
        &\lesssim
        \frac{
        \qty(A(\nin)\alpha(\nin)\nin^2)^{A(\nin)\alpha(\nin)\nin^2}
        }{
        \qty(A(\nin)\alpha(\nin)\nin^2-\alpha(\nin)\nin)^{
        A(\nin)\alpha(\nin)\nin^2-\alpha(\nin)\nin}
        \qty(\alpha(\nin)\nin)^{\alpha(\nin)\nin}
        } \\
        &=
        \frac{
        \qty(A(\nin)\alpha(\nin)\nin^2)^{A(\nin)\alpha(\nin)\nin^2}
        \qty(A(\nin)\alpha(\nin)\nin^2-\alpha(\nin)\nin)^{\alpha(\nin)\nin}
        }{
        \qty(A(\nin)\alpha(\nin)\nin^2-\alpha(\nin)\nin)^{
        A(\nin)\alpha(\nin)\nin^2}
        \qty(\alpha(\nin)\nin)^{\alpha(\nin)\nin}
        } \\
        &=
        \qty(
        \frac{A(\nin)\alpha(\nin)\nin^2}
        {A(\nin)\alpha(\nin)\nin^2-\alpha(\nin)\nin}
        )^{A(\nin)\alpha(\nin)\nin^2}
        \qty(
        \frac{A(\nin)\alpha(\nin)\nin^2-\alpha(\nin)\nin}
        {\alpha(\nin)\nin}
        )^{\alpha(\nin)\nin}.
    \end{split}
    \label{app_eq:binomial_expansion}
\end{align}
Here $\lesssim$ means that we ignore subexponential factors, which do not affect the sufficient scaling of $\ein'$ chosen below.
The first factor in \Cref{app_eq:binomial_expansion} is bounded as
\begin{align}
    \qty(
    \frac{A(\nin)\alpha(\nin)\nin^2}
    {A(\nin)\alpha(\nin)\nin^2-\alpha(\nin)\nin}
    )^{A(\nin)\alpha(\nin)\nin^2}
    &=
    \qty(
    1+\frac{1}{A(\nin)\nin-1}
    )^{A(\nin)\alpha(\nin)\nin^2} \notag\\
    &\leq e^{\alpha(\nin)\nin}
    =
    e^{\din}.
\end{align}
The second factor is
\begin{align}
    \qty(\frac{A(\nin)\alpha(\nin)\nin^2-\alpha(\nin)\nin}
    {\alpha(\nin)\nin})^{\alpha(\nin)\nin}=\qty(A(\nin)\nin-1)^{\din}.
\end{align}
Therefore,
\begin{align}
    \binom{n'}{\din}
    \lesssim
    \qty(e[A(\nin)\nin-1])^{\din}.
    \label{app_eq:binomial_bound_refined}
\end{align}
In particular, the following coarser bound is sufficient for our purpose:
\begin{align}
    \binom{n'}{\din}
    \leq
    \qty(A(\nin)\nin\din)^{\din},
    \label{app_eq:binomial_bound_coarse}
\end{align}
up to irrelevant constant and subexponential factors.

Therefore, we choose
\begin{align}\label{app_eq:condition_of_input}
    \ein'
    =
    \qty(\frac{1}{A(\nin)\nin\din})^{\din}.
\end{align}
Indeed, combining \Cref{app_eq:binomial_bound_coarse} with \Cref{app_eq:condition_of_input}, we obtain
\begin{align}
    \binom{n'}{\din}(\ein')^{\din}\leq\qty(A(\nin)\nin\din)^{\din}\qty(\frac{1}{A(\nin)\nin\din})^{\din^2}\xrightarrow[\nin\to\infty]{}0,
\end{align}
because $\din\to\infty$ and $A(\nin)\nin\din\to\infty$. 
Although \Cref{app_eq:binomial_bound_coarse} is not tight, the slack is useful because we only need a sufficient scaling of $\ein'$ that guarantees both vanishing output error and high post-selection success probability.

We now include the cost of preparing intermediate magic states with the error rate in \Cref{app_eq:condition_of_input}. 
Suppose that the pre-distillation step has an overhead exponent $\gpred$. 
Then the number of original noisy magic states required to prepare one intermediate magic state with error $\ein'$ scales as
\begin{align}
    C_p=O\qty(\log^{\gpred}\frac{1}{\ein'}).
\end{align}
Substituting \Cref{app_eq:condition_of_input}, we get
\begin{align}
    C_p
    &=
    O\qty(\log^{\gpred}\qty[
    \qty(A(\nin)\nin\din)^{\din}]) \notag\\
    &=
    O\qty(\din^{\gpred}\log^{\gpred}\qty(A(\nin)\nin\din)).
    \label{app_eq:pre_distillation_cost}
\end{align}
Thus, although $\ein'$ must vanish with the code parameters, this requirement only contributes a polylogarithmic overhead in $1/\ein'$.

Taking the pre-distillation cost into account, the total number of original noisy magic states is
\begin{align}
    n
    =
    C_p n'.
\end{align}
Using \Cref{app_eq:nprime_asymptotic} and \Cref{app_eq:pre_distillation_cost}, we obtain
\begin{align}
    n = O\qty(A(\nin)\nin \din^{\gpred+1}\log^{\gpred}\qty(A(\nin)\nin\din)).
    \label{app_eq:total_input_cost}
\end{align}

Then, we introduce an exponent $t$ characterizing the leading order of the distillation rate.
Specifically, define
\begin{align}
    \xi
    \coloneqq
    \frac{a_n^t}{n}.
\end{align}
Here, $\xi$ should be understood as the leading-order coefficient in the scaling relation between the total input size $n$ and the output size $a_n$. 

If $\xi$ remains finite and bounded away from zero for some $t$, then the total input cost scales at most as $a_n^t$ up to constant factors. 
Thus, it is enough to show that the parameters can be chosen so that $t\to1^+$ without making $\xi$ vanish in the asymptotic limit.

Since $a_n=A(\nin)\kin(\nin)=A(\nin)\nin(1+o(1))$, \Cref{app_eq:total_input_cost} gives, up to constant factors,
\begin{align}
    \xi=\frac{[A(\nin)\nin]^{t-1}}{\din^{\gpred+1}\log^{\gpred}\qty(A(\nin)\nin\din)}(1+o(1)).
    \label{app_eq:rate_culc}
\end{align}
Solving \Cref{app_eq:rate_culc} for $t$, we obtain
\begin{align}
    t=1+\frac{(\gpred+1)\log\din+\gpred\log\log\qty(A(\nin)\nin\din)+\log \xi}{\log\qty(A(\nin)\nin)}
    +o(1).
    \label{app_eq:solve_t}
\end{align}

It remains to show that the second term in \Cref{app_eq:solve_t} can be made $o(1)$. 
By the outer code construction discussed in \Cref{def:outercode}, for a family of inner codes with parameters $\codepar{\nin,\kin(\nin),\din(\nin)}$, the factor $A(\nin)$ can be chosen to grow polynomially in $\kin(\nin)^{\din(\nin)}$. 
Since $\kin(\nin)=\nin(1-o(1))$, this implies that, for an appropriate choice of the outer code family,
\begin{align}
    \log\qty(A(\nin)\nin)=  \Omega\qty(\din\log\nin).
\end{align}
On the other hand, the numerator in the second term of \Cref{app_eq:solve_t} is
\begin{align}
    O\qty(\log\din+\log\log\qty(A(\nin)\nin\din)).
\end{align}
Under the above choice of $A(\nin)$, this is at most logarithmic in $\din\log\nin$, whereas the denominator grows as $\Omega(\din\log\nin)$. 
Therefore, the second term in \Cref{app_eq:solve_t} is $o(1)$, and hence
\begin{align}
    t\xrightarrow[\nin\to\infty]{}1^+.
\end{align}
This shows that the total input cost can be made $a_n^{1+o(1)}$, or equivalently, that the distillation rate can be made arbitrarily close to linear in the asymptotic limit.

We finally relate this almost-linear scaling to the rate condition stated in \Cref{thm:asymptotic_sublinear}. 
From \Cref{app_eq:rate_culc}, setting $t=1$ gives the explicit rate scaling
\begin{align}
    \frac{a_n}{n}=O\qty(\frac{1}{\din^{\gpred+1}
    \log^{\gpred}\qty(A(\nin)\nin\din)}),
\end{align}
up to constant and subleading factors. 
Since the pre-distillation protocol used here has $\gpred>1$ and the inner code distance satisfies $\din=\omega(1)$, the denominator grows faster than $\log n$, where $n$ denotes the total number of input magic states. 
Therefore,
\begin{align}
    \frac{a_n}{n}=o\qty(\frac{1}{\log n}).
\end{align}

\section{Finite size distillation }\label{app:finite_size}
To demonstrate \Cref{thm:asymptotic_sublinear} in this paper, we compare the protocol we proposed with the asymptotically constant-overhead protocol of \cite{Wills2024-im} and a basic small-size $(15\to 1)$ protocol of \cite{Bravyi2012-us} in a finite regime.  
In this section, we present the specific parameters of the protocols used in this comparison.
We begin by explaining the common framework and the associated rules used for the performance comparison of the protocols. The target state to be distilled is $\ket{T} $. The states $\ket{H}$ and $\ket{T}$ are related by $\ket{T}=HS\ket{H}$, and since both $H$ and $S$ are Clifford gates, it follows that these states can be converted into each other at a unit rate. Furthermore, for all protocols under consideration, the pre-distillation step is performed using the $15\to 1$ protocol~\cite{Bravyi2012-us}.

In this paper, the comparison is carried out under the assumption that the infidelity of the initial noisy magic states is fixed to $\ein = 0.1$. The protocols are then compared in terms of the yield of magic state distillation and the error suppression order, namely, the resulting infidelity after distillation $\eout$.

In the \Cref{subsection:15to1}, we describe the orthodox $15 \to 1$ protocol.
In the following \Cref{subsection:COMS}, we explain the constant magic state distillation protocol.

\subsection{Repetition of small protocol ($15\to1$)}\label{subsection:15to1}
In this section, we describe the performance of magic state distillation based on the Reed–Muller code \cite{Bravyi2012-us}.
The Reed–Muller code is a $\codepar{15,1,3}$ code and admits a transversal $T$ gate $\qty(T^{\otimes 15}=\Bar{T})$ here, $\Bar{T}$ means a logical $T$ gate. 
That is, applying $\Bar{T}$ to the encoded state $\ket{\Bar{+}}$ in the Reed–Muller code, one can implement a magic state distillation protocol with input magic count $15$, output magic count $1$, and error suppression $35\ein^{3}$. Here, the coefficient $35$ denotes the total number of undetectable weight-three errors.
From this, if magic state distillation based on the Reed–Muller code is iterated $p$ times, the input–output ratio of magic states $R_{15\to1}^p$ and the infidelity of the output state $\eout^{15\to1}_p$ can be expressed as follows.
\begin{equation}
    R_{15\to1}^p=\qty(\frac{1}{15})^p, \quad \varepsilon^{15\to1}_p=35^{\qty(\frac{3^{p}-1}{2})}\cdot \ein^{3^{p}}.
\end{equation}
Moreover, the success probability at the $b$-th stage of pre-distillation $P_{\rm succ}^{\rm pre}$ can be written as follows.
\begin{align}
    P_{\rm succ}^{\rm pre}=\prod_{p=1}^{b}\qty(1-\varepsilon_{p-1}^{15\to 1})^{15},
\end{align}
where, $\varepsilon_{p-1}$ is output error rate about $(p-1)$-th stage, and $\varepsilon_0\coloneqq\ein=0.1$.
Thus effective distillation rate of pre-distillation $R_{\rm 15\to1}^{\rm eff}$ is
\begin{align}\label{app_eq:15to1}
    R_{\rm 15\to1}^{p\rm, eff}:= P_{\rm succ}^{\rm pre}\cdot R_{15\to1}^p
\end{align}
The purple points in \Cref{fig:comparison_with_quantum algebraic geometry} correspond to plots of \Cref{app_eq:15to1}.

\subsection{Protocol of constant overhead magic state distillation}\label{subsection:COMS}
Magic state distillation with asymptotically constant overhead has recently been  achieved~\cite{Wills2024-im}.
 Their strategy does not use post-selection however rather corrects the error. Thus, we denote the quantum algebraic geometry code as $\codepar{N,K,D,r_d}_{\rm qAG}$ where $r$ is the decoding radius. First, we describe the parameters of the quantum algebraic geometry code used for comparison. This is based on concrete parameters in Section 5.3 in~\cite{Wills2024-im}. In this section, we do not dive into the details of algebraic geometry codes; instead, we focus only on the $\codepar{n,k,d}$ codes employed for comparison and the corresponding distillation protocols.
\begin{align}\label{eq:parameter_of_qag}
        \left\lfloor \frac{118}{4}\left(33\times 32^{i-1}-34\times 32^{\frac{i-1}{2}}+1\right)\right\rfloor -1 \geq \; &N_i \geq \left\lfloor \frac{114}{4}\left(33\times 32^{i-1}-34\times 32^{\frac{i-1}{2}}+1\right)\right\rfloor-3,\\
        &K_i=\left\lfloor\frac{5}{4}\left(33\times 32^{i-1}-34\times 32^{\frac{i-1}{2}}+1\right)\right\rfloor,\\
        &D_i\geq\left\lfloor\frac{5}{4}\left(33\times 32^{i-1}-34\times 32^{\frac{i-1}{2}}+1\right)\right\rfloor+1,\\
        &r_i=\left\lfloor\frac{1}{8}\left(33\times 32^{i-1}-34\times 32^{\frac{i-1}{2}}+1\right)\right\rfloor,
    \end{align}
    and for all $i\in\{4,6,8,\ldots\}$,
    \begin{align}
         \left\lfloor \frac{118}{4}\left(33 \times 32^{i-1}-561\times 32^{\frac{i}{2}-1}+1\right)\right\rfloor -1 \geq \; &N_i \geq \left\lfloor \frac{114}{4}\left(33 \times 32^{i-1}-561\times 32^{\frac{i}{2}-1}+1\right)\right\rfloor -3 ,\\
        &K_i=\left\lfloor \frac{5}{4}\left(33 \times 32^{i-1}-561\times 32^{\frac{i}{2}-1}+1\right)\right\rfloor,\\
        &D_i\geq\left\lfloor \frac{5}{4}\left(33 \times 32^{i-1}-561\times 32^{\frac{i}{2}-1}+1\right)\right\rfloor+1,\\
        &r_i=\left\lfloor \frac{1}{8}\left(33 \times 32^{i-1}-561\times 32^{\frac{i}{2}-1}+1\right)\right\rfloor.
    \end{align}
    Given that their protocol is of constant rate, the comparison reduces to evaluating the input–output rate associated with the smallest code construction. For comparison purposes, we consider the most advantageous instance in the above expression: specifically, the case where $N$ is minimized. In their protocol, to use error correction instead of error detection and post-selection, note that the error suppression order depends $r_i$ not $D_i$ ($\ein\to O(\ein^{r_{i}})$). Thus, the minimum construction of the quantum algebraic geometry code is $\codepar{932093,40881,d,4088}_{1024}$, here, $\codepar{\cdot}_{q}$ means quantum error-correcting code on $q$-dimensional qudit system.
    
Second, we describe the transition chart using the quantum algebraic geometry code for the distillation of the state $\ket{T}$.
\begin{align}
    \ket{T}\overset{\text{pre-distill}}{\longrightarrow}\ket{T}\overset{4 /1}{\longrightarrow}\ket{CCZ}\overset{70 / 1}{\longrightarrow}\ket{U_s}\overset{\text{distill}\hspace{1mm} N / K}{\longrightarrow}\ket{U_s}\overset{1 / 1}{\longrightarrow}\ket{CCZ}\otimes \underset{1 \text{catalyst}}{\ket{T}} \overset{1/2}{\longrightarrow}\ket{T}
\end{align}
Here, we use the 15 to 1 protocol using Reed Muller code~\cite{Bravyi2012-us} for the pre-distill step for $\ket{T}$. And conversion of $4\ket{T}\to\ket{CCZ}$ is along with~\cite{Beverland2020-hm}, and $\ket{CCZ}\otimes\ket{T}\to\ket{T}$ is along with in~\cite{Gidney2019-ag}. Note that the $\ket{T}$ state for the catalyst must be reduced to an error rate equivalent to the last target state in the pre-distill phase. When their method considers the distillation of $\ket{T}$ states with constant overhead, they limit the number of $\ket{T}$ states prepared as a catalyst to one and use the method of using all of them. If the initial error rate is $\ein$ and the target error rate is $\eoutqag$, it costs about $15^{l_c}$ (where $l_c$ is the smallest integer that satisfies the following equation) to distill the $\ket{T}$ states for the catalyst.
\begin{align}\label{error_reduction_of_RM}
    \eoutqag > 35^{(\frac{3^{l_c}-1}{2})}\ein^{3^{l_c}}.
\end{align}
The constant-overhead protocol based on quantum algebraic geometry codes requires the input magic states to be below a certain threshold error rate.  In this context, the threshold refers to a constant upper bound on the input error rate such that the output error decreases with increasing code size.  This threshold condition is needed to determine how accurately the input magic states must be prepared before applying the constant overhead  protocol.  A lower bound on the threshold can be obtained from the following leading-order estimate: 
\begin{align}\label{eq:error_bound}
    \eoutqag&\leq\binom{N}{r_d+1}(C_{\mathrm{conv}}\ein)^{r_d+1}\\
    &\leq 2^{N h\left(\frac{r_d+1}{N}\right)}(C_{\mathrm{conv}}\ein)^{r_d+1}\\
    &=\left(\frac{\ein}{1/\qty(C_{\mathrm{conv}}2^{\frac{N}{r_d+1}h\left(\frac{r_d+1}{N}\right)})}\right)^{r_d+1},
\end{align}
where $h(x)\coloneqq-x\log_2(x)-(1-x)\log_2(1-x)$ is the binary entropy function,  $C_{\mathrm{conv}}$ is the conversion rate of $\ket{CCZ}\to \ket{U_s}$, and they confirmed the existence of $C_{\mathrm{conv}}=70$, and the second inequality follows from $\binom{N}{K}\leq 2^{N h(K/N)}$. Then, due to $r_d=\Theta(N)$, there exists a threshold of error rate of input magic states;
\begin{align}
    \varepsilon_\mathrm{th}\geq\liminf_{N\to\infty}\frac{1}{C_{\mathrm{conv}}2^{\frac{N}{r_d+1}h\left(\frac{r_d+1}{N}\right)}}>0.
\end{align}
By substituting the numerical values obtained from ~\Cref{eq:parameter_of_qag}, we obtain the following.
\begin{align}
    \varepsilon_{\mathrm{th}}&\geq \frac{1}{C_{\mathrm{conv}}\cdot 2^{\frac{N}{r_d+1}h(\frac{r_d+1}{N})}}\\
    &=\frac{1}{70\cdot 2^{\frac{76}{3}h(\frac{3}{76})}}\\
    &=2.14\times 10^{-4},\\
    \frac{r_d+1}{N}&\simeq \frac{\frac{1}{8}+1}{\frac{114}{4}}=\frac{3}{76},
\end{align}

If the initial error rate is set as 0.1, then at least $l_c=3$ is needed to fall below this threshold. Consider a connection to the quantum algebraic geometry code with $l_c=3$. The error rate after pre-distillation is $1.18\times 10^{-7}$ by substituting the condition into the expression \Cref{error_reduction_of_RM}. In other words, the error rate after distillation by the quantum algebraic geometry code performed is as follows.
\begin{align}
    \eoutqag &\leq \qty(\frac{1.18\times 10^{-7}}{2.14\times 10^{-4}})^{(r_d+1)}\\
    &=\qty(5.5\times 10^{-4})^{(r_d+1)}.
\end{align}
Therefore, one $\ket{T}$ state for a catalyst must be pre-distilled with the Reed-Muller code a minimum of $l_c$ times, satisfying the following conditions.
\begin{align}
    35^{\frac{3^{l_c}-1}{2}}0.1^{3^{l_c}}\leq \qty(5.5\times 10^{-4})^{(r_d+1)}.
\end{align}
Now recalling that $r_d=N/228$, we can rewrite.
\begin{align}
    \frac{3^{l_c}-1}{2}\log 35 +3^{l_c} \log 0.1 \leq \qty(\frac{N}{228}+1)\log(5.5\times 10^{-4}).
\end{align}
i.e. $l_c=O(\log N)$. From the above, the distillation rate $R^{\ket{T}}_{\rm qAG}$ is 
\begin{align}
    R^{\ket{T}}_{\rm qAG}&=\frac{2K+1}{15^3 \cdot 280N +15^{l_c}}\\
    &\simeq \frac{5N}{57\cdot 945000N}\\
    &=\frac{1}{10773171}=9.28\times 10^{-8}.\label{eq:rate_of_qag}
\end{align}

\subsection{Sublinear protocol in finite size distillation}\label{app_subsection:finite_ours}

Next, we consider the degree of error rate reduction achieved by our protocol for comparison. Since our protocol has a sublinear rate, the distillation rate is monotonically decreasing with respect to $n$.  
That is, among the constructions for which the distillation rate of our protocol, calculated by ~\Cref{eq:distillation-rate_ours}, exceeds the value given in~\Cref{eq:rate_of_qag}, one should choose the one that minimizes $\eout$. To ensure a fair comparison, in the pre-distillation stage of our protocol we use a protocol for this competition based on the $15 \to 1$ protocol using the Reed-Muller code.
\begin{align}
    \ket{T}\overset{\text{pre-distill}}{\longrightarrow}\ket{T}\overset{1}{\longrightarrow}\ket{H}\overset{\text{distill}}{\longrightarrow}\ket{H}\overset{1}{\longrightarrow}\ket{T}
\end{align}
That is, the effective rate of our protocol $R^{\rm eff}_{\rm f}$ can be written as follows.
\begin{align}
    R^{\rm eff}_{\rm f}&=P_{\rm succ}^{\rm pre}\qty(1-\ein')^{n'}\frac{a_n}{n} \\
    &=\frac{P_{\rm succ}^{\rm pre}}{15^{l_p}}\cdot (1-\ein')^{n'}\frac{A(\nin)\nin\qty(1-\frac{
    \din}{\nin})}{A(\nin)\nin \din},\label{eq:app_eff_of_finite}
\end{align}
where $l_p$ is the number of repetition times of pre-distillation. Moreover, the first part of \Cref{eq:app_eff_of_finite} corresponds to the effective distillation rate in pre-distillation given by \Cref{app_eq:15to1}, while the latter part corresponds to the effective distillation rate of the sublinear protocol, namely the product of the success probability of post-selection and $\frac{a_n}{n'}$. In particular, the final fractional term in \Cref{eq:app_eff_of_finite} is obtained by substituting \Cref{eq:app_inputcount} and \Cref{eq:app_a_n} into $\frac{a_n}{n'}$.

Here, note that we assume $A(\nin) = \kin^\din$.
In this case,
\begin{align}
    \ein' =\qty(\frac{1}{\kin^\din  \din})^\din.
\end{align}
where, $\ein'$ denotes the error rate that must be achieved by the pre-distillation step. And this condition is from \Cref{app_eq:condition_of_input}. 
Moreover, in our calculations we take $\din = \log \nin$ as a function in $\omega(1)$.
That is, the parameters of the inner code we consider are given by
\begin{align}
    \left\llbracket \nin, \kin=\left\lfloor\nin\qty(1-2h\qty(\frac{\lfloor\log \nin\rfloor}{\nin}))\right\rfloor, \din=\lfloor\log \nin\rfloor\right\rrbracket.
\end{align}
At this point, we search for the smallest $\eout$ such that $R_{\rm qAG} < R^{\rm eff}_{\rm f}$ and $\din$ is odd.  
As a result, we obtain $\codepar{\nin=8104,\kin=8002,\din=9}$, $l_p=5$ and in this case, $\eout=10^{-353}$ is achieved. 
The distillation rate in this case is $R^{\rm eff}_{\rm f}=2.1\times 10^{-6}$. It should be noted here that this value does not correspond to the optimal choice of parameters but merely demonstrates that such parameters do exist.  
Taking this fact together with the observation that our rate decreases monotonically with respect to $-\log \eout$, we find that our magic state distillation protocol achieving a sublinear rate remains superior over a fairly wide intermediate regime.

In this sense, the main magic state distillation protocol presented in this paper can be regarded as an intermediate scheme between small-size protocols and protocols with constant overhead.
Finally, we summarize in a \Cref{tab:forfig} the distillation rates and achieved error rates corresponding to the blue points plotted as our scheme in \Cref{fig:comparison_with_quantum algebraic geometry}. The quantum error-correcting code used in this simulation has parameters $\codepar{8104, 8002, 9}$.

\begin{table}[h]
    \centering
    \begin{tabular}{c|c|c}
       $(p,r)$  &  distillation rate& $\eout$ \\ \hline
         (3,1)& $6.6\times 10^{-6}$ & $10^{-15}$\\
        (4,1) & $2.5\times 10^{-7}$ & $10^{-353}$
    \end{tabular}
    \caption{Coordinates of the points corresponding to the blue points in \Cref{fig:comparison_with_quantum algebraic geometry}.
}
    \label{tab:forfig}
\end{table}

\begin{figure}
    \centering
    \includegraphics[width=1\linewidth]{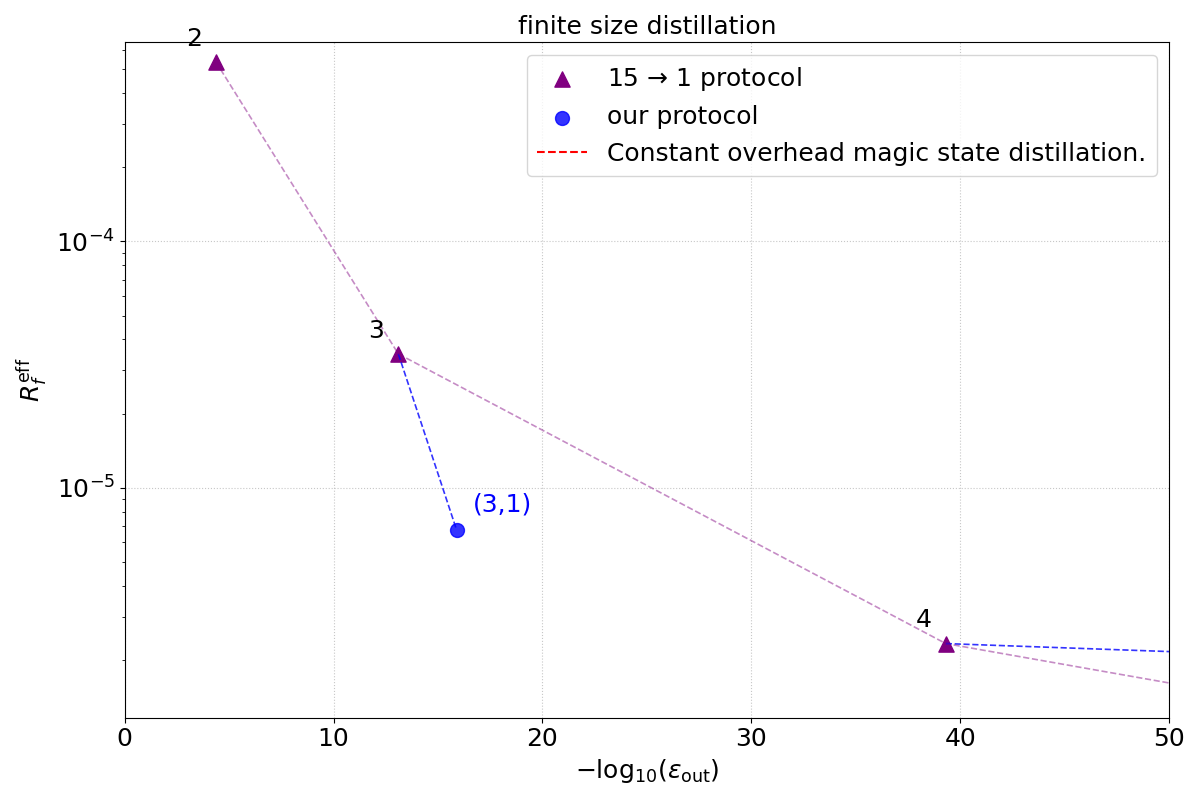}
    \caption{Magnified view of the large-error regime in \Cref{fig:comparison_with_quantum algebraic geometry}. The purple points represent the $15\to1$ protocol, while the blue points represent our protocol. The blue point $(3,1)$ corresponds to performing distillation using $\codepar{149,117,5}$ after three rounds of pre-distillation. This point achieves less error reduction than the purple point $(4)$, namely four rounds of pre-distillation, but is able to maintain a higher rate.
}
    \label{app_fig:zoom_fig2}
\end{figure}

\end{document}